\documentclass[%
 reprint,
 amsmath,amssymb,
 aps,
]{revtex4-1}

\usepackage{graphicx}
\usepackage{dcolumn}
\usepackage{bm}
\usepackage{mathrsfs}
\usepackage{ulem}
\usepackage{color}

\usepackage{csquotes}

\begin{document}

\preprint{APS/123-QED}

\title{Exciton-polaritons in flatland:\\ Controlling flatband properties in a Lieb lattice}

\author{Tristan H. Harder$^1$}
 \email{tristan.harder@physik.uni-wuerzburg.de}
\author{Oleg A. Egorov$^2$}
\author{Johannes Beierlein$^1$}%
\author{Philipp Gagel$^1$}
\author{Johannes Michl$^1$}
\author{Monika Emmerling$^1$}%
\author{Christian Schneider$^1$}%
\author{Ulf Peschel$^2$}
\author{Sven H\"ofling$^{1,3}$}     
\author{Sebastian Klembt$^1$}       
 \email{sebastian.klembt@physik.uni-wuerzburg.de}

\affiliation{$^1$Technische Physik, Wilhelm-Conrad-R\"ontgen-Research Center for Complex Material Systems, and  W\"urzburg-Dresden Cluster of Excellence ct.qmat, Universit\"at W\"urzburg, Am Hubland, D-97074 W\"urzburg,
Germany}
\affiliation{$^2$Institute of Condensed Matter Theory and Solid State Optics, Abbe Center of Photonics, Friedrich-Schiller-Universit\"at Jena, D-07743, Germany}
\affiliation{$^3$SUPA, School of Physics and Astronomy, University of St Andrews, St Andrews
KY16 9SS, United Kingdom}

\date{\today}

\begin{abstract}
In recent years, novel two-dimensional materials such as graphene, bismuthene and transition-metal dichalcogenides have attracted considerable interest due to their unique physical properties. A range of physical effects can be transferred to the realms of photonics by creating artificial photonic lattices emulating these two-dimensional materials. Here, exciton-polaritons in semiconductor microcavities offer an exciting opportunity to study a part-light, part-matter quantum fluid of light in a complex lattice potential. In this paper, we study exciton-polaritons in a two-dimensional Lieb lattice of buried optical traps. The $S$ and $P_{xy}$ photonic orbitals of such a Lieb lattice give rise to the formation of two flatbands which are of greatest interest for the distortion-free storage of compact localized states. By using a well controlled etch-and-overgrowth technique, we manage to control the trapping as well as the site couplings with great precision. This allows us to spectroscopically monitor the flatness of the flatbands across the full Brillouin zone. Furthermore, we demonstrate experimentally that these flatbands can be directly populated by condensation under non-resonant laser excitation. Finally, using this advanced device approach we demonstrate resonant and deterministic excitation of flatband modes in transmission geometry. Our findings establish the exciton-polariton systems as a highly controllable, optical many-body system to study flatband effects and for distortion-free storage of compact localized states.

\end{abstract}

\pacs{Valid PACS appear here}
\maketitle



Flatbands are a fascinating class of completely dispersionless bands, emerging from symmetry considerations and phase frustrations in a range of tight-binding Hamiltonians of periodic lattices. Consequently, the energy $E(k)$ of a flatband is independent of the Bloch state momentum $k$. Such bands have been predicted \cite{Lieb1989} and subsequently used for the theoretical description of physical phenomena ranging from itinerant ferromagnetism \cite{Lieb1989, Tasaki1992}, to fractional quantum Hall states \cite{Tsui1982} and topological flatband phases \cite{Wang2011, Goldman2011}. Here, the so-called Lieb lattice, arising from a slightly altered square lattice (see Figs.~1(a)-(d)), is of greatest theoretical interest as it features  distinct Dirac-cone dispersions at the $M$-points as well as a dispersionless flatband cutting through the Dirac points. With an overall interest and improved mastery of synthetic quantum matter \cite{Polini2013}, this geometry has been used to realize flatband effects experimentally in systems such as cold atoms \cite{Taie2015, Ozawa2017}, electronic surface states \cite{Kalff2016,Slot2017}, photonics \cite{Guzman2014,Mukherjee2015,Vincencio2015,Real2017,Leykam2017a,Leykam2017b}, exciton-polaritons in 1D \cite{Baboux2016,Goblot2019} and 2D lattices \cite{Klembt2017,Whittaker2018}.\\
Microcavity exciton-polaritons (polaritons) \cite{Weisbuch1992} are hybrid light-matter particles arising from the strong coupling between excitons and microcavity photons. Their strong nonlinearities inherited from the matter part in combination with their accessibility by angular-resolved photo- or electroluminescence (PL and EL) spectroscopy granted by the light part put them in the focus of fundamental research devoted to macroscopic quantum phases of exciton-polaritons \cite{Carusotto2013}. Most notably, they have shown the ability to undergo bosonic condensation \cite{Kasprzak2006,Schneider2013}, superfluid behavior \cite{Amo2009} and the formation of a topological Chern insulator mode \cite{Klembt2018}. 
The appeal of polariton physics in search of novel many-body phenomena stems from the high degree of interactions in combination with advanced technological control over the semiconductor microcavity structures hosting polaritons \cite{Schneider2017}. The most evident approach towards polariton confinement is to etch micropillar resonators with diameters of several micrometers into planar microcavities. Due to the optical confinement, discrete modes occur. By designing these pillars with overlap and thus coupling between them, two-dimensional lattices forming band structures can be fabricated and studied \cite{Bayer1998, Jacqmin2014}. Downsides of this method include a large etch-surface leading to surface defects, a limited coupling range due to overlap requirements, as well as a lack of control over the depth of the confinement potential.\\ 
\begin{figure*}[ht!]
  \includegraphics[width=0.80\textwidth]{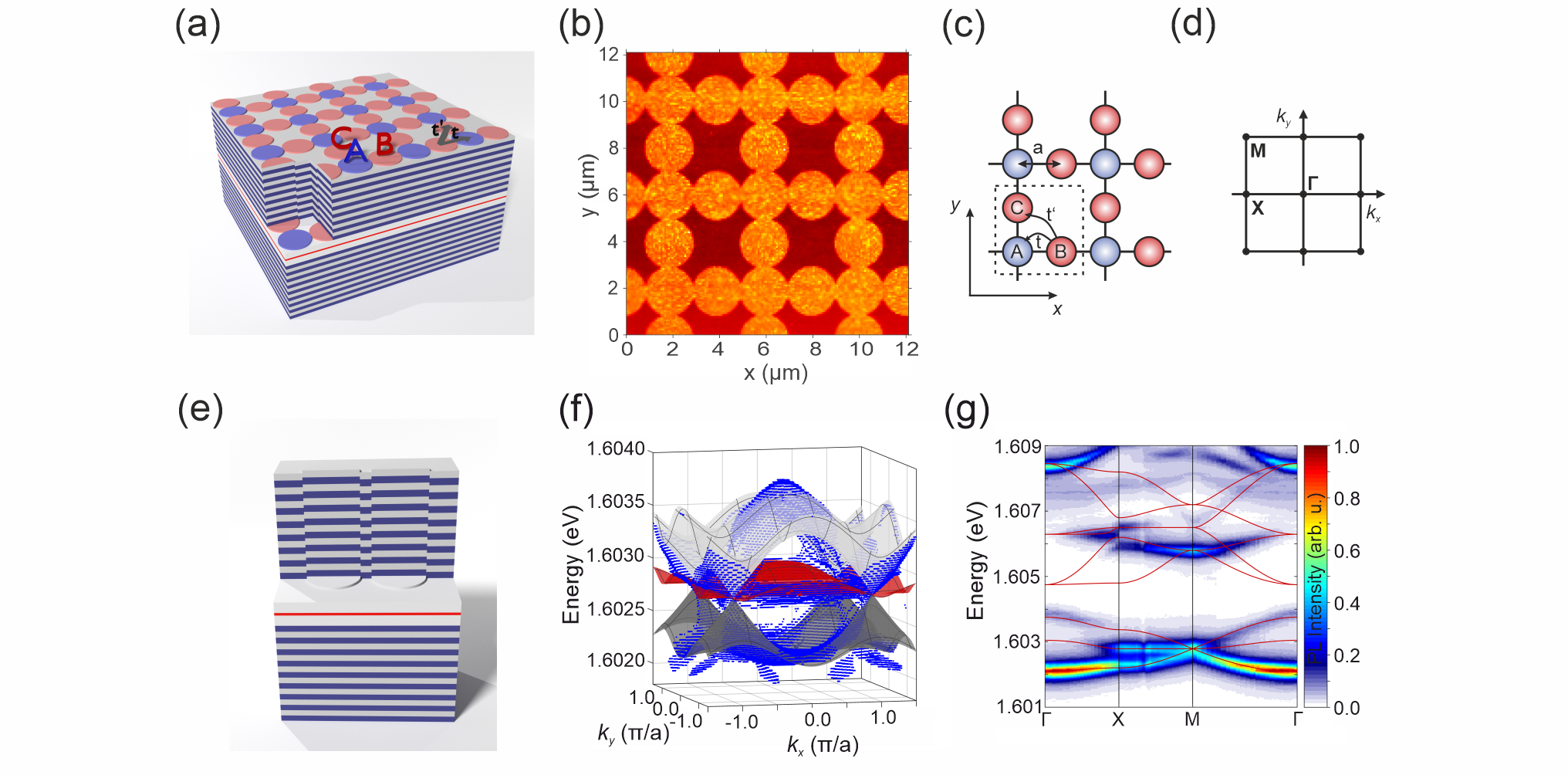}
  \caption{
  (a) Schematic drawing of the investigated etch-and-overgrowth Lieb lattice. A, B and C denote the three sites of the lattice unit cell, \textit{t} and \textit{t}' describe the nearest and next-nearest neighbor coupling, respectively. (b) Atomic force microscope image of the etched spacer layer. The Lieb lattice sites are elevated by $\sim$5~nm.(c) Real space and (d) first Brillouin zone representation of the Lieb lattice with $\Gamma$, $X$ and $M$ denoting the high symmetry points and $a$ being the center-to-center distance. (e) Zoom on two coupled nearest neighbor sites in the etch-and-overgrowth geometry. Pronounced coupling and mode hybridization is possible even for traps that do not overlap. (f) Measured Fourier space energy-resolved photoluminescence of the $S$-band of the investigated Lieb lattice in $k_x$-direction, scanning in $k_y$-direction. The blue data points are extracted from the measured dispersions, accurately revealing the four Dirac cones at the $M$-points. In red (flatband) and gray (dispersive bands) a tight-binding model is plotted, agreeing very well with the measured data. (g) Photoluminescence measurements of the $S$- and $P$-bands for a Lieb lattice array of polariton traps in the reduced BZ representation. The $S$- and $P$-flatbands are visible at around 1.603\,eV and 1.606\,eV, respectively.}
\end{figure*}
A way to overcome these limitations is the so-called etch-and-overgrowth (EnO) technique to confine polaritons \cite{Daif2006,Winkler2015,Kuznetsov2018}. Here, the molecular beam epitaxial growth is interrupted after finishing the cavity layers. The trapping potential is created by patterning and subsequent wet etching of a spacer layer located on top of the actual cavity (Fig.~1(a)). Fig. 1(b) shows an atomic force microscope image of traps in a Lieb lattice geometry with a trap height of approximately 5\,nm. In Figs. 1(c) and (d), the unit cell of the real space lattice and the first Brillouin zone with its high-symmetry points, respectively, are presented. Finally, the top distributed Bragg reflector (DBR) is epitaxially grown. In this approach, the trapping potential is created by a modal red shift induced through a local elongation of the cavity. As the confinement potential can be finely tuned by changing the etching depth, the coupled traps can be geometrically separated (see Fig. 1(e)) and still form a band structure.\\
While a range of semiconductor materials have been shown to host microcavity polaritons, the mature technological control over the GaAs-based system facilitates it as an ideal candidate for sophisticated microcavity lattice structures. The polariton lattice investigated in this paper is a vertically emitting microcavity with 32 Al$_{0.20}$Ga$_{0.80}$As/AlAs mirror pairs in the top DBR, 37 Al$_{0.20}$Ga$_{0.80}$As/AlAs mirror pairs in the bottom DBR,  and two sets of 4 GaAs quantum wells (QWs) embedded in the field antinode of the  AlAs $\lambda/2$-wavelength-thick cavity and in the first mirror pair of the bottom DBR. The exciton energy is $E_X=1.610\,$eV. We characterized the sample with white light reflectance measurements at 10 K by using the spatial variation of the cavity resonance across the wafer and extracted a vacuum Rabi splitting of 2$\hbar \Omega_R$ = $11.5$\,meV. The EnO-traps lead to a confinement of $E_{conf}\sim6.2\,$meV. The quality factor of the cavity was extracted from PL measurements on a far red-detuned area of the wafer, yielding a value of Q\,$\sim$\,7,500. All measurements have been performed at a moderately negative detuning $\delta \sim -3.9\,$meV.\\
\begin{figure*}
  \includegraphics[width=0.65\linewidth]{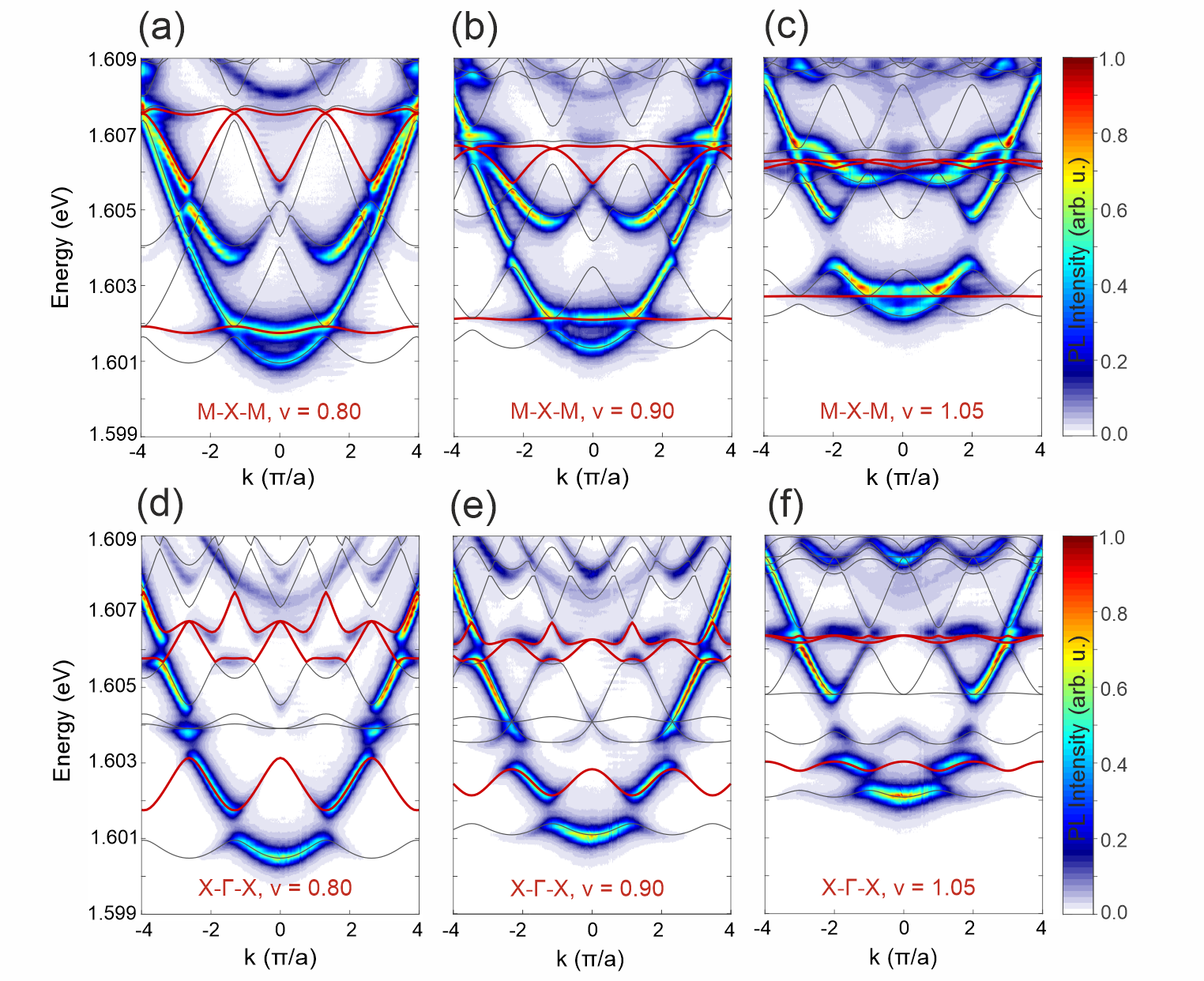}
  \caption{Dispersion measurements below threshold for reduced trap distances of v=0.80, v=0.90 and v=1.05 along the (a)-(c) $M-X-M$- and (d)-(f) $X-\Gamma -X$ directions, respectively. Particularly in $X-\Gamma-X$ direction, the next-nearest neighbor coupling \textit{t}' leads to the strongest deviation from the tight-binding model and thus to a curvature of the flatbands. A full Bloch mode calculation describes accurately the band structure (gray lines) as well as the respective flatbands (red lines).}
\end{figure*}
Optical characterization is performed by  momentum-resolved PL spectroscopy. For the PL measurements, we excite our sample with a continuous wave (cw) Ti:sapphire laser tuned to the reflectance minimum of the first high-energy Bragg mode at 1.658~eV. The luminescence is collected using a Fourier spectroscopy setup with a Cherny-Turner spectrometer and a Peltier-cooled 1024x1024 px CCD camera operating at -75 $^{\circ}$C. By motorized scanning of the last imaging lens, both in real space and Fourier space imaging configuration, we can collect the full dispersion information in $k_x$- and $k_y$-directions, with $k_{||}=\sqrt{k_x^2+k_y^2}$ being the in-plane wave vector, as well as perform optical tomographies in real space ($E(x,y)$). All experiments have been carried out in a liquid helium flow cryostat at  $T=4\,$K.\\
In the Lieb lattice, the flatband occurs as a consequence of destructive interference in the lattice coupling. Starting from a tight-binding Hamiltonian in real space, one can obtain the Fourier space  Hamiltonian of the $S$-band 
\begin{equation}
\mathscr{H}_{TB} ({\bf{k}} ) =
   \begin{pmatrix}
   0 & t+t e^{-ik_x} & t+te^{-ik_y}  \\
   t+t e^{ik_x} & 0 & t^\prime f({\bf{k}} ) \\
   t+t e^{ik_y} & t^\prime f^* ({\bf{k}} )  & 0  \\
   \end{pmatrix},
\end{equation}
with $t$ and $t^\prime$  being the inter-site and next-nearest neighbor couplings, respectively, and $f({\bf{k}} )\equiv( e^{ik_x}+ e^{-ik_y}+1)$.  $ {\bf{k}} =\left\{k_{x},k_{y}\right\}  $  is the reciprocal lattice vector, defined in the first Brillouin zone. In the simplest case of nearest-neighbor coupling only ($t^\prime=0$), one directly obtains the dispersion relation
\begin{equation}
E( {\bf{k}} )= 0;  \pm 2 t\sqrt{ \cos^2(k_x/2) + \cos^2(k_y/2) }
\end{equation}
featuring a flatband that is, in this simplest case, perfectly flat. However, in any realistic systems a next-nearest neighbor coupling $t'$ (see Fig. 1(c)) cannot be avoided.\\
\begin{figure}
  \includegraphics[width=0.85\linewidth]{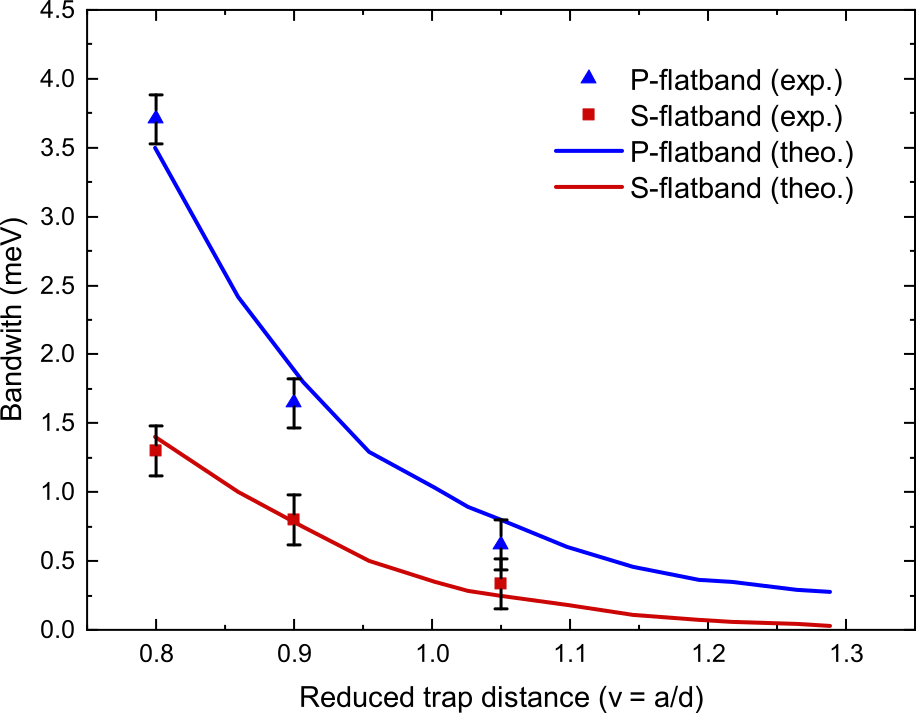}
  \caption{Measured spectral bandwidth of the $S$- and $P$-flatbands (symbols) compared to the theoretically predicted bandwidths, extracted from the calculations  (solid lines). For a reduced trap distance of $v>1.2$ the bandwidth is expected to be well within the linewidth of $\sim$ 300\,$\mu$eV in the linear regime.}
\end{figure}
While the typical real space signature of the flatband, which is occupation occuring only on the $B$- and $C$-sites, referred to as Lieb diamond, has been described in a range of papers \cite{Mukherjee2015,Vincencio2015,Klembt2017,Whittaker2018}, full tomographic measurements on the flatband in Fourier space have been missing so far. Fig. 1(f) depicts a tomographic characterization of the polariton Lieb-lattice in the linear regime, using non-resonant laser excitation. The Fourier space energy-resolved photoluminescence of the investigated lattice is imaged in the $k_x$-direction and scanned in the $k_y$-direction. The blue data points are extracted from the measured dispersions, accurately revealing the four Dirac cones at the $M$ points as well as the $S$-flatband. 
The corresponding tight-binding model with a nearest neighbor coupling of \textit{t}~=~360~$\mu$eV and a next-nearest neighbor coupling of \textit{t}'~=~80~$\mu$eV is plotted in gray with the flatband in red and agrees well with the experimental data. Fig. 1(g) depicts the $S$- and $P$-bands for a Lieb lattice array of polariton traps in the reduced Brillouin zone representation.\\
In order to systematically study the influence of next-nearest neighbor coupling $t'$ on the flatness of the flatband as well as deformation of the modes of separated mesas, we now vary the coupling conditions in the lattices. The investigated structures have a trap diameter $d=2\, \mu m$ and a reduced trap distance $v=a/d = 0.80; 0.90; 1.05$, where $a$ is the center-to-center distance of adjacent traps (Fig. 1(c)). Therefore, a value $v<1$ corresponds to overlapping traps, $v=1$ indicates touching traps, and $v>1$ denotes geometrically separated traps.\\
Since the separations between the mesas in the lattice are comparable or even smaller then their sizes, we expect substantial deformation of the mode profiles of separate mesas. Thus the tight binding approach is not valid any more. In order to realistically describe the polariton Lieb lattice in an EnO structure, we determine the energy-momentum band structure of the Lieb lattices using a full description of the Bloch modes taking into account all relevant system parameters. For this aim we solve the following eigenvalue problem for the energy $ E ({\bf{k}}_b) $ of the Bloch mode with the Bloch vector $ {\bf{k}}_b =\left\{k_{bx},k_{by}\right\}  $   
\begin{equation}\label{eq:eigenvalue}
E({\bf{k}}_b) \left\{ {\begin{array}{*{20}{c}}	{{p_b}({\bf{r}},{{\bf{k}}_b})}\\
	{{e_b}({\bf{r}},{{\bf{k}}_b})}	\end{array}} \right\} = \hat L({\bf{k}}_b) \left\{{\begin{array}{*{20}{c}}	{{p_b}({\bf{r}},{{\bf{k}}_b})}\\
	{{e_b}({\bf{r}},{{\bf{k}}_b})}	\end{array}} \right\},
\end{equation}
where the functions $ {p_b}({\bf{r}},{{\bf{k}}_b}) $ and $ {e_b}({\bf{r}},{{\bf{k}}_b}) $ describe the amplitude distributions of the photonic and excitonic component of the Bloch modes in real space defined in the plane of the microcavity $ {\bf{r}}=\left\{x,y\right\} $. The main matrix in Eq.~(\ref{eq:eigenvalue}), describing the single-particle coupled states of excitons and photons, is given by the expression
 \begin{equation}\label{}
 \hat L({\bf{k}}_b) =\hbar \left( {\begin{array}{*{20}{c}}
 	{ \omega _C^0 + V({\bf{r}}) - \frac{{{\hbar}}  {{\left( {\vec \nabla _ \bot ^{} + i{{\bf{k}}_b}} \right)}^2}   }{{2{m_C}}}\quad \quad \quad \Omega \quad }\\
 	{\quad \Omega \quad \quad \quad \quad \quad {\kern 1pt} \quad \omega _E^0 - \frac{{{\hbar}} {{\left( {\vec \nabla _ \bot ^{} + i{{\bf{k}}_b}} \right)}^2}    }{{2{m_E}}}}
 	\end{array}} \right).   \nonumber
 \end{equation}
In the model above, the quantities $ \omega _C^0 $ and $ \omega _E^0 $ represent the energies of bare photons and excitons, respectively. The photon-exciton coupling strength is given by the parameter $ \hbar \Omega $ which defines the Rabi splitting for the microtraps. Here, $ m_C = 34.3 \times 10^{-6}m_e$ is the effective photon mass in the planar region and $ m_e$ denotes the free electron mass. The effective mass of excitons is $ m_X \approx 10^{5}m_C$. An external photonic potential $V({\bf{r}})$ is defined within the unit cell of the Lieb structure  compound of mesas.

In Fig. 2, the results of a systematic variation of the reduced trap distance on the $S$- and $P$-flatband are presented. Here, Fig. 2(a)-(c) show the PL spectra in $M$-$X$-$M$ and Fig. 2(d)-(f) in $X$-$\Gamma$-$X$ direction for v=0.80 to v=1.05, respectively. While the $M$-$X$-$M$ spectra exhibit flat bands within the linewidth throughout the trap distance variation, the dispersions in $X$-$\Gamma$-$X$ direction feature a distinct curvature of the flatbands. As expected from the full Bloch mode calculation, we are able to reduce the energy bandwidth of the $S$- and $P$-flatbands below $\Delta E_{S,P} = 500\,\mu$eV by increasing the reduced trap distance to v=1.05, thus decreasing the influence of next-nearest neighbors. The rather shallow confinement potential of 6.2\,meV allows for a physical separation of the traps by 5$\%$ (and more) of their diameter while keeping a lattice band structure. The full Bloch mode calculations (gray, flatbands red) in Fig. 2(a)-(f) are in excellent agreement with the spectroscopic data. The respective bandwidths of the $S$- and $P$-flatbands are plotted against the reduced trap diameter $v$ in Fig. 3, where the symbols represent the experimental findings for the S- and P-flatbands and the lines result from the Bloch mode calculations. We can extrapolate that for v$\sim$1.20 the flatband bandwidth will be well below the respective linewidth of approximately 300\,$\mu$eV of the system, where at the same time a distinct band structure formation can still be well expected.

\begin{figure}
  \includegraphics[width=1.00\linewidth]{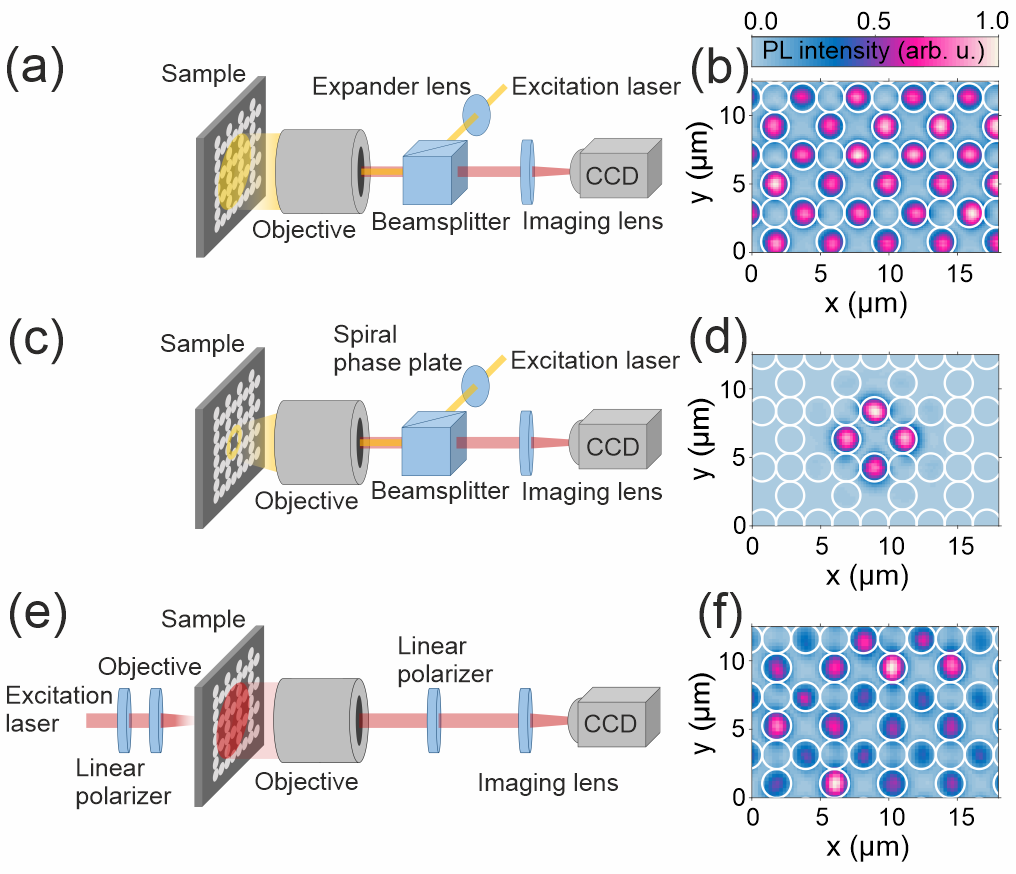}
  \caption{Overview of the different spectroscopic methods used to populate the hybrid light-matter flatband states, using the $S$-flatband as an example. (a,b) Non-resonant cw laser excitation with a large spot, leading to a uniformly occupied $S$-flatband condensate accross several tens of unit cells. (c,d) Non-resonant cw laser excitation of a polariton condensate in a single compact flatband site (Lieb diamond) using a spiral phase plate. (e,f) Resonant cw laser excitation of the $S$-flatband in transmission, using a backside polished sample and polarization optics to suppress the excitation laser. (b,d,f) Real space images of the respective PL emission from the Lieb flatband. All methods show the distinct diamond shaped, real space signature of the Lieb flatband.}
\end{figure}

In order to demonstrate the potential of EnO microcavity designs to host hybrid light-matter flatband states, we now use a range of optical techniques to populate flatband states with polaritons. In Fig. 4(a), the excitation scheme for non-resonant cw laser excitation with a large spot covering a multitude of lattice sites is illustrated. An appropriately chosen exciton-photon detuning of $\delta \sim -3.9\,$meV allows for polariton condensation into the $S$-flatband, characteristically represented by the diamond-shaped mode pattern in Fig. 4(b) (see \cite{Klembt2017,Whittaker2018} for comparison). Corresponding spectra and threshold characteristics can be found in the Supplement. Having established the excitation of a large flatband condensate, we continue by demonstrating the excitation of a compact localized condensate in a nearly flat band as theoretically proposed and described by Sun et al. \cite{Sun2018}. For this purpose, we use a spiral phase plate to control the phase and intensity of a ring shaped Laguerre-Gaussian beam profile with a diameter of $d\sim 4.0\,\mu$m centered at the Lieb lattice diamond (B and C sites). When increasing the excitation power, we observe polariton lasing from a single compact localized state (CLS). CLSs represent a key element of localized information in a flatband system \cite{Maimaiti2017} and can be easiliy adressed spectroscopically in our polariton lattice, as depicted in Figs. 4(c) and (d). Non-resonant excitation schemes might allow for the realization of collective flatband lasing \cite{Longhi2019}, potentially even using electrical injection \cite{Suchomel2018}.\\
Having established non-resonant control of the flatband states, finally, we make use of another technological advantage of EnO microcavity structures to deterministically address polariton flatband states by \textit{resonant} laser excitation.  We use a cw laser with a spot size of $\sim30\mu$m in transmission geometry by exciting through the polished sample backside. The EnO microcavity used is very similar to the one investigated so far, but features three In$_{0.04}$Ga$_{0.96}$As QWs instead of GaAs QWs and is characterised by a Rabi splitting of $\sim$4.5~meV. The exciton energy of these quantum wells is lower than the band gap of the GaAs substrate, which hence becomes transparent to the excitation laser at E$_{cw}=1.4711$eV. To allow resonant excitation in the transmission scheme presented in Fig. 4(e), the backside of the sample is polished using a lapping plate. A detailed description of the sample preparation for transmission measurements can be found in the Supplement. The cw laser is carefully prepared to be linearly polarized and subsequently filtered in the detection using cross-polarization. As etching of micropillars is not required when defining the lattice potential with the EnO technique, both the polished backside as well as the front surface of the sample are smooth and scattering of the laser is minimized. Fig. 4(f) highlights again the well known real space flatband signature and verifies the selective and controlled occupation of the desired flatband state. The overall intensity fluctuation is directly related to the specific laser excitation and transmission properties of the sample. Particularly, the resonant excitation of polaritons in flatband states opens entirely new ways to study interactions \cite{Kuno2019}, scattering dynamics \cite{ Gneiting2018}, and topological effects \cite{Sigurdsson2017,Li2018,Zhang2019}. Using spatial light modulators, a highly controllable excitation of CLS of any taylorable shape becomes conceivable. \\
In conclusion, we have successfully designed and demonstrated the use of etch-and-overgrowth traps in a Lieb lattice geometry to create polariton flatband states. Using advanced tomography techniques we are able to directly image these flatbands. Furthermore, by optimizing the highly accessible coupling parameters we are able to to flatten the flatbands such that the spectral bandwidth of the flatbands approaches the linewidth of the flatband itself. We have demonstrated that the EnO fabrication technique allows for a highly controllable population of polaritonic flatbands and compact localized states via non-resonant excitation and condensation. Finally, we have been able to show the first experimental results on fully resonant excitation of any modes in polariton lattices, specifically demonstrated for a lattice flatband mode.
This work is an important step towards new polaritonic platforms with properties and functionalities involving topology, gain, and interactions \cite{Ozawa2019}.



The W\"{u}rzburg side acknowledge support by the W\"urzburg-Dresden Cluster of Excellence on Complexity and Topology in Quantum Matter - ct.qmat (EXC 2147). S.K., J.B., U.P. and O.A.E. acknowledge support by the German Research Foundation (DFG) within project KL2431/2-1. S.H. is furthermore grateful for support within the EPSRC ”Hybrid Polaritonics” Grant (EP/M025330/1). T.H.H. and S.H. acknowledge funding by the doctoral training program \enquote{Elitenetzwerk Bayern}. T.H.H. acknowledges support by the German Academic Scholarship Foundation.

\end{document}